\documentclass[letterpaper, 10 pt, conference]{ieeeconf}  

\IEEEoverridecommandlockouts                              

\usepackage{cite}
\usepackage{amsmath,amssymb,amsfonts}
\usepackage{algorithmic}
\usepackage{graphicx}

\usepackage{algorithm}
\usepackage{subcaption}
\usepackage{siunitx}
\usepackage{multirow}

\newcommand{\xx}{\mathbf{x}}

\newcommand{\uu}{\mathbf{u}}

\newcommand{\XX}{\mathbf{X}}
\newcommand{\UU}{\mathbf{U}}
\newcommand{\LL}{\mathbf{\Lambda}}

\newcommand{\lam}{\boldsymbol{\lambda}}
\newcommand{\pppi}{\boldsymbol{\Pi}}
\newcommand{\tx}{\mathrm}

\DeclareMathOperator*{\argmin}{arg\,min}
\DeclareMathOperator*{\indicator}{I}
\DeclareMathOperator*{\shift}{shift}

\AfterEndEnvironment{Definition}{\noindent\ignorespaces}

\title{\LARGE \bf
Interaction-Aware Trajectory Prediction and Planning in Dense Highway Traffic using Distributed Model Predictive Control
}

\author{Erik Börve, Nikolce Murgovski, and Leo Laine
\thanks{This work was partially supported by the Wallenberg AI, Autonomous Systems and Software Program (WASP) funded by the Knut and Alice Wallenberg Foundation.}
\thanks{Erik Börve and Nikolce Murgovski are with the Department of Electrical Engineering, Chalmers University of Technology, 412 96 Göteborg, Sweden {\tt\footnotesize\{borerik, nikolce.murgovski\}@chalmers.se}}%
\thanks{Leo Laine is with the Department of Mechanics and Maritime Science, Chalmers University of Technology, 412 96 Göteborg, Sweden {\tt\footnotesize leo.laine@chalmers.se}}%
}

\begin{document}

\maketitle
\thispagestyle{empty}
\pagestyle{empty}

\begin{abstract}
In this paper we treat optimal trajectory planning for an autonomous vehicle (AV) operating in dense traffic, where vehicles closely interact with each other. To tackle this problem, we present a novel framework that couples trajectory prediction and planning in multi-agent environments, using distributed model predictive control. A demonstration of our framework is presented in simulation, employing a trajectory planner using non-linear model predictive control. We analyze performance and convergence of our framework, subject to different prediction errors. The results indicate that the obtained locally optimal solutions are improved, compared with decoupled prediction and planning.
\end{abstract}

\section{Introduction} \label{sec:introduction}

\begin{figure}
    \centering
    \includegraphics[width = 0.5 \textwidth,trim={2cm 0 0 0},clip]{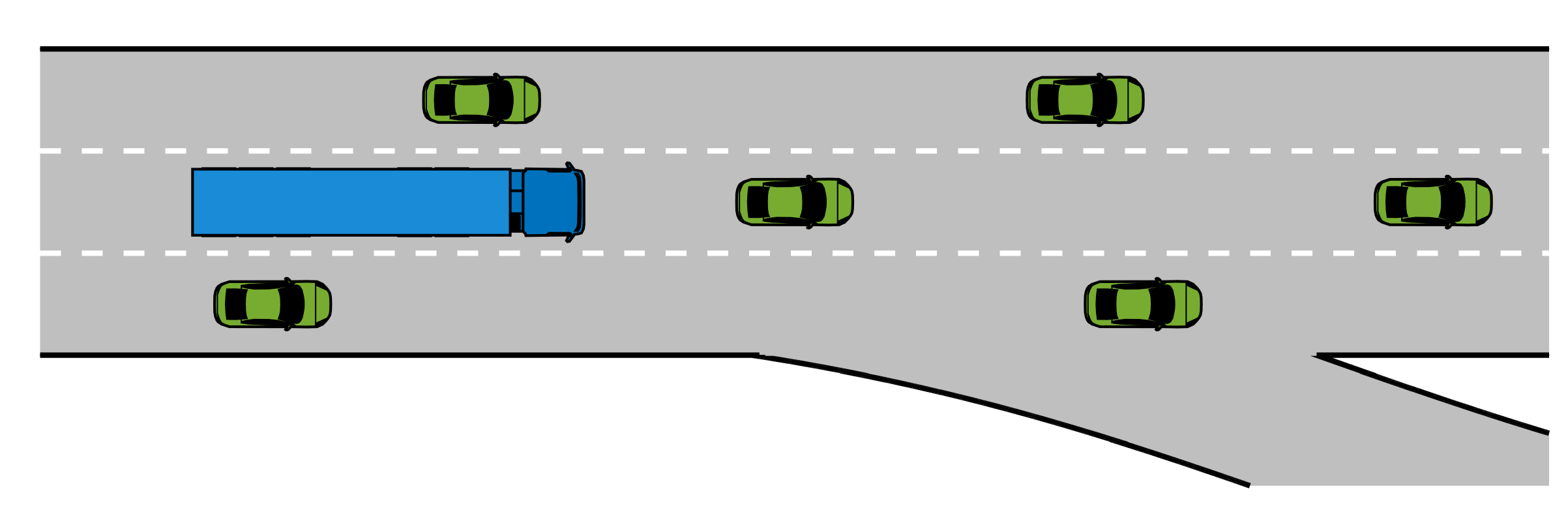}
    \caption{Forced lane-change scenario for an HVC (blue) approaching an exit ramp on a multi-lane highway.}
    \label{fig:exScenario}
\end{figure}

 Considering the vast amount of existing traffic situations, trajectory planning in safety-critical traffic scenarios remains an active research topic. In this paper we consider one such scenario: highway driving maneuvers in \textit{dense} traffic, where inter-vehicle distances can be smaller than the vehicles length. These maneuvers depend on complex driver-to-driver interactions, which pose challenges as the possibility for communication between vehicles is limited in practice \cite{trafficInteractions}. Most commonly, drivers utilize the turn signal and adjust the relative distances to show intention. Further, maneuvers become additionally challenging when including heavy vehicle combinations (HVCs), due to e.g., limited visibility, and larger vehicle dimensions \cite{nilssonSimulator} (see, e.g., a forced lane change in Fig.~\ref{fig:exScenario}).
 
Navigating in traffic can be considered as path planning with moving obstacles. Prior work has addressed this problem with Model Predictive Control (MPC), grid/tree-searching methods, and learning-based methods, see \cite{trajectoryReview} and sources therein for a detailed review. The MPC approaches \cite{overtakeMPC,laneMergeSOTA,karlssonExitMPC} have received attention due to the ability to enforce rigorous constraints on the dynamics of the own vehicle (henceforth referred to as the ego-vehicle) and safety with respect to avoiding collisions with other vehicles. In this context, MPC iteratively solves a constrained optimization problem based on current measurements of the environment and computes control actions for a finite time/space window, referred to as the horizon. However, without vehicle-to-vehicle communication, MPC requires a prediction of the surrounding traffic over the horizon \cite{trajectoryReview}. Since humans (drivers) do not abide by Newtonian laws, predicting their trajectories is non-trivial and especially challenging when considering crowded environments with many interpersonal dependencies. The prediction of such trajectories has recently been addressed by learning-based predictors \cite{trajectron++}, which may also account for interactions among surrounding vehicles. The predicted trajectories can then be used to formulate collision avoidance constraints for an MPC-based trajectory planner, showing promising results in simulations of dense traffic scenarios \cite{laneMergeSOTA}.
In such interactive scenarios, there exists an inherent dependence between the planned MPC trajectory and the predicted trajectories of other road users. Hence, by decoupling the predictor and the planner, the predictions become less accurate and the MPC is not able to 
account for interactions with other vehicles. In current practical applications, interactive lane change maneuvers for AVs can become excessively defensive or even be avoided \cite{planningAVs,nilssonSimulator}.

To address this problem, we propose to couple the predictor with the planning controller. To the best of our knowledge, we propose a novel method based on Distributed Model Predictive Control (DMPC) and demonstrate the possibilities with a highway-autopilot controller for an HVC that considers interactions. The proposed method can in principle be combined with any predictor, possibly non-differentiable, that can be conditioned on a given trajectory, such as that proposed in \cite{trajectron++}. Hence, our main contributions are as follows:
\begin{enumerate}
    \item We propose a novel method for combining trajectory prediction and planning by utilizing DMPC. 
    \item We demonstrate our approach for a novel problem, employing non-linear MPC for an HVC in dense, and interaction-dependant traffic.
    \item We provide an empirical analysis of the effect of prediction errors on the MPC performance and the convergence of our proposed method.
\end{enumerate}
A simulation environment together with a model-based predictor was developed for the purpose of this paper.


\section{Distributed Optimal Control}\label{sec:DMPC_theory}
MPC has been used in a wide array of multi-agent scenarios to optimize performance metrics and ensure safety-critical constraints. One implementation of MPC is a centralized approach, which considers a single controller that optimizes the combined objective of all agents over the horizon with respect to all states $\xx$ and control variables $\uu$, most often subject to some constraints. In network control problems, agents with local objectives, local constraints, and coupled inequality constraints, are often encountered. Such problems can be solved centrally, but their computational complexity typically scales poorly with an increasing number of agents \cite{centralizedVSdist}.
One approach to address this is to decompose the central problem using DMPC. The network control problem can instead be expressed for each agent $i$ over a discrete horizon $k = 0, \dots, N$ as,
\begin{equation} \label{eq:P_dist}
    \begin{split}
        \min_{\xx_i,\uu_i} & f_{\tx{f},i}(\mathbf{x}_i(N)) + \sum_{k=0}^{N-1} f_i(\mathbf{x}_i(k),\mathbf{u}_i(k))\\
        \text{s.t. } &\mathbf{x}_i(k+1) = g_i(\mathbf{x}_i(k),\mathbf{u}_i(k)),\;k = 0,\dots, N-1\\
        &h_i(\mathbf{x}(k),\mathbf{u}(k)) \leq 0, \quad \quad \quad \quad \; k = 0,\dots, N-1 \\
        &\mathbf{x}_i(0) = \mathbf{x}_{0,i}
    \end{split}
\end{equation}
where $f_{\tx{f},i}$ and $f_i$ describe the target and running cost of each agent, respectively, $g_i$ describes the dynamics, $h_i$ describes the coupled constraints, $\mathbf{x}_i$ and $\mathbf{u}_i$ are states and control inputs, respectively, and $\mathbf{x}_{0,i}$ are initial states. The coupling among the agents occurs in the inequality constraint $h_i$, which is a function of the states and control inputs of all the agents. A solution for the central problem is then obtained by overhead communication between all $M$ agents. Rawlings et. al. \cite{rawlings} describes ``Non-cooperative'' DMPC, 
where communication is restricted to the solution of each agent's respective optimization problem. In this setting, agents aim to minimize their own objective, treating other agents predicted trajectories $(\mathbf{x}_i, \mathbf{u}_i) ^p$ as known disturbances. Each agent's beliefs of the other agents' trajectories are updated using a convex combination of their current optimal solution $(\mathbf{x}_i, \mathbf{u}_i) ^\star$ and the current iterate $(\mathbf{x}_i, \mathbf{u}_i) ^p$ as,
\begin{equation} \label{eq:convex_step}
    \begin{split}
    \mathbf{x}_i^{p+1} & \leftarrow w_i \mathbf{x}_i^\star + (1-w_i) \mathbf{x}_i^p\\
    \mathbf{u}_i^{p+1} &\leftarrow w_i \mathbf{u}_i^\star + (1-w_i) \mathbf{u}_i^p.
    \end{split}
\end{equation}

The step size $w_i$ represents each agent's belief in their own current optimal solution where $0 < w_i < 1$. In practice, the convex step is iterated until convergence within a certain tolerance.

Do however note that for our problem we only control the ego-vehicle, while the trajectories of surrounding vehicles are estimated by a predictor. Considering non-Newtonian human drivers, the learned control policies of the surrounding vehicles could be non-differentiable, or even obtained by black-box approaches, and closed-form expression for their $f_i$, $g_i$ or $h_i$ functions may not be available. Yet we show in this paper that it may still be possible to implement a DMPC strategy using a gradient-based method for the ego-vehicle controller.


\section{Vehicle Modelling}
This section provides a model of the ego vehicle and the surrounding traffic used in simulations.

\subsection{Ego-Vehicle Model}
The ego-vehicle motion is modeled using a kinematic bicycle model, extended with a trailer, as
\begin{equation} \label{eq:ego_dynamics}
    \Dot{\mathbf{x}}_\tx{e} = \begin{bmatrix}
        \Dot{p}_{x}\\
        \Dot{p}_{y}\\
        \Dot{v}_{x}\\
        \Dot{\theta_1}\\
        \Dot{\theta_2}
        \end{bmatrix} = \begin{bmatrix}
            v_{x}\\
            v_{x} \tan{\theta_1}\\
            a_{v} \cos{\theta_1}\\
            v_{x}\frac{\tan{\delta}}{\ell_1 \cos{\theta_1}}\\
            v_{x}\frac{\sin{\left(\theta_1 - \theta_2\right)}}{\ell_2 \cos{\theta_1}}
        \end{bmatrix}
, \quad
    \mathbf{u_\tx{e}} = \begin{bmatrix}
        \delta\\
        a_{v}
    \end{bmatrix}.
\end{equation}
Here, $\xx_\tx{e}$ are states that include x- and y-position at the tractor-trailer joint in the inertial frame $p_{x},p_{y}$, longitudinal velocity at the tractor-trailer joint in the inertial frame $v_{x}$ and the tractor and trailer angles with respect to the inertial frame $\theta_1,\theta_2$. The vehicle states are controlled by the steering angle $\delta$ and the longitudinal acceleration $a_v$ at the tractor-trailer joint expressed in the body frame. The wheelbase of the tractor and trailer are represented by $\ell_1,\ell_2$, respectively. For simulation and optimization, the model is discretized as,
\begin{equation}
    \xx_\tx{e}(k+1) = g_\tx{e}(\xx_\tx{e}(k), \mathbf{u}_\tx{e}(k)).
\end{equation}


\subsection{Traffic Model} \label{sec:traffic_model}
The surrounding vehicles are modeled using a kinematic bicycle model. The state $\xx_i$ and control vector $\uu_i$ of a vehicle $i$ are described as for the ego-vehicle but excluding the trailer state $\theta_2$. The longitudinal acceleration of each vehicle $a_i$ is calculated based on tracking its own reference velocity and maintaining a safe distance to leading vehicles in the same lane. Additionally, vehicles can consider accelerating or decelerating to allow a vehicle in an adjacent lane to merge into their own, the extent of which is determined by how cooperative the vehicle is \cite{brito2022learning
}. Such a traffic model can be described with a general function as,
\begin{equation}
        \xx_i(k+1) = g_i(\xx(k),a_i(k), \boldsymbol{\phi})
\end{equation}
where $\boldsymbol{\phi}$ are model parameters and  $\xx$ are the states of all surrounding vehicles. The steering angle $\delta$ of surrounding vehicles has been considered zero in the studied traffic scenarios, as focus is placed on whether surrounding vehicles can yield a sufficient gap for the ego vehicle to fit into by accelerating or decelerating. However, it is straightforward to apply the proposed method to scenarios where surrounding vehicle may also change lanes.

\section{Ego-Vehicle Problem Formulation} \label{sec:ego_problem}
The trajectory planning problem, even for a single vehicle, is a mixed-integer program due to the need of choosing a target lane which is inherently an integer decision. Instead of solving a single mixed-integer problem, it was proposed in \cite{overtakeMPC,karlssonExitMPC} to solve three MPCs, each with a different target lane, i.e.,
keep current lane (no lane change), change to the left lane, and change to the right lane, noted as ${j \in \{\tx{nc},\tx{lc},\tx{rc}\}}$, respectively. A decision manager compares the costs of the optimal MPC solutions and returns the safe optimal choice, e.g., driving forward, and a desired optimal choice, e.g., a forced lane change. 

Compared to \cite{overtakeMPC,karlssonExitMPC}, we additionally consider a scenario with blocking traffic where the ego-vehicle is not able to immediately initiate a lane change, but it instead cautiously approaches adjacent vehicles to gauge their cooperativeness. 

\subsection{Collision Avoidance Constraints}
As the collision avoidance constraints are based on estimations of trajectories $(\hat{\xx}_i^j,\hat{\uu}_i^j)$ of surrounding vehicle $i$ for each controller $j$, deviations from the true trajectories can yield infeasible problems in edge cases. Hence, we introduce slack variables $\lam_\tx{e}^j \in \mathbb{R}^{N \times M_c^j}$ to relax the collision avoidance constraints, where $\mathbb{M}^j_\tx{v}$ represents a subset of the surrounding vehicles, with size $M_c^j$. These receive a substantial penalty in the objective to remove the possibility of collisions in a practical sense but persistent feasibility is no longer rigorously established.

\begin{figure*}[t!]
    \vspace{3mm}
    \centering
    \begin{subfigure}{0.32\textwidth}
    \includegraphics[width = 1\textwidth,trim={0cm 0.6cm 0cm 0},clip]{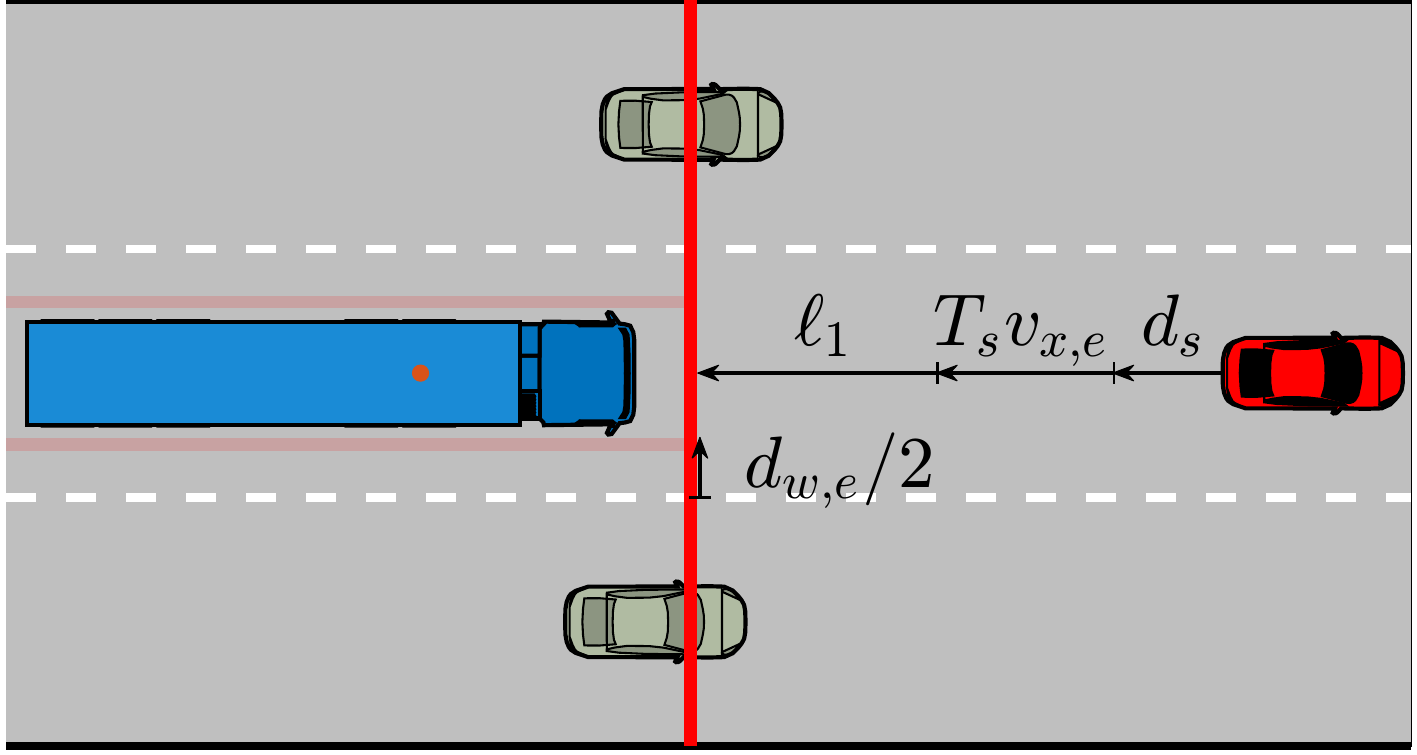}
    \caption{Collision avoidance constraints for the lane keeping controller.}
    \label{fig:constraint_keeplane}
    \end{subfigure}
    \centering
    \begin{subfigure}{0.32\textwidth}
    \includegraphics[width = 1\textwidth,trim={0 0.6cm 0 0},clip]{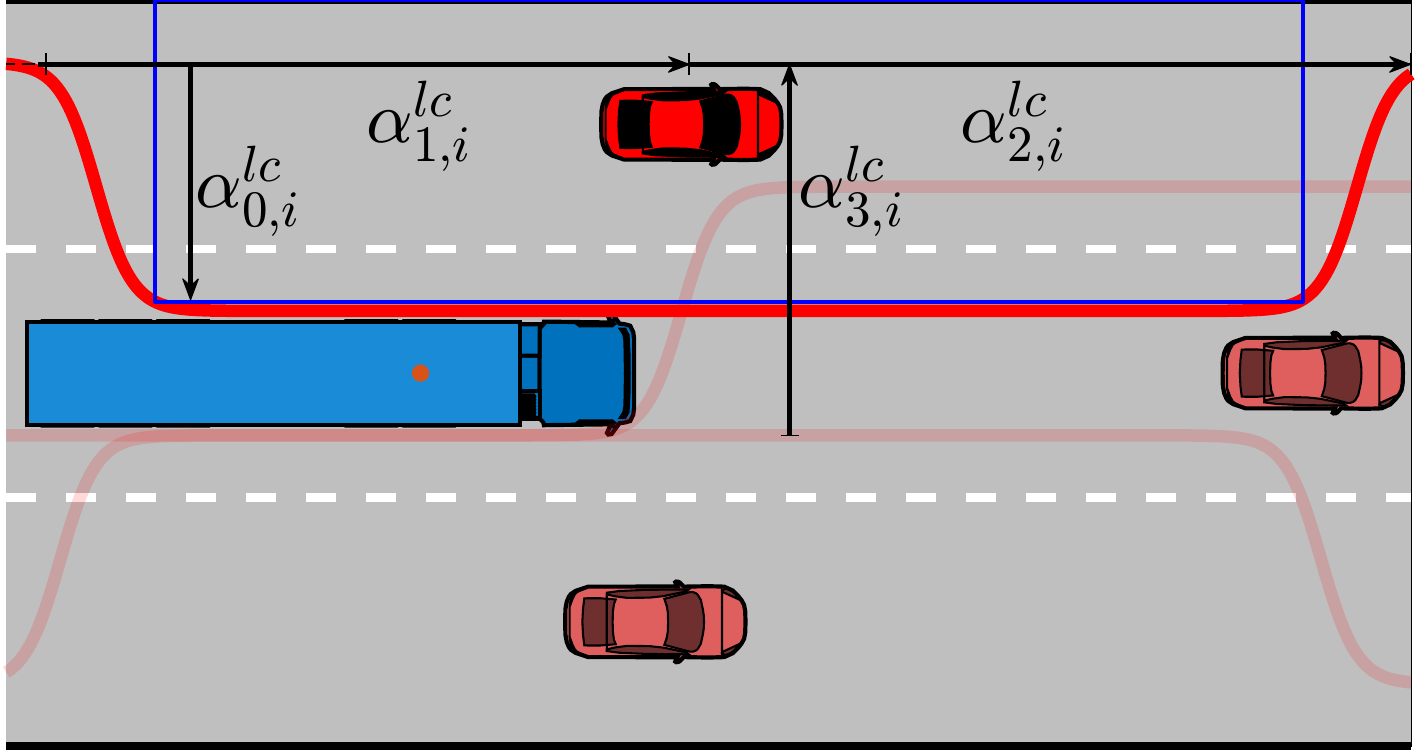}
    \caption{Collision avoidance constraints for the left lane change controller.}
    \label{fig:constraint_changeleft}
    \end{subfigure}
    \centering
    \begin{subfigure}{0.32\textwidth}
    \includegraphics[width = 1\textwidth,trim={0 0.6cm 0 0},clip]{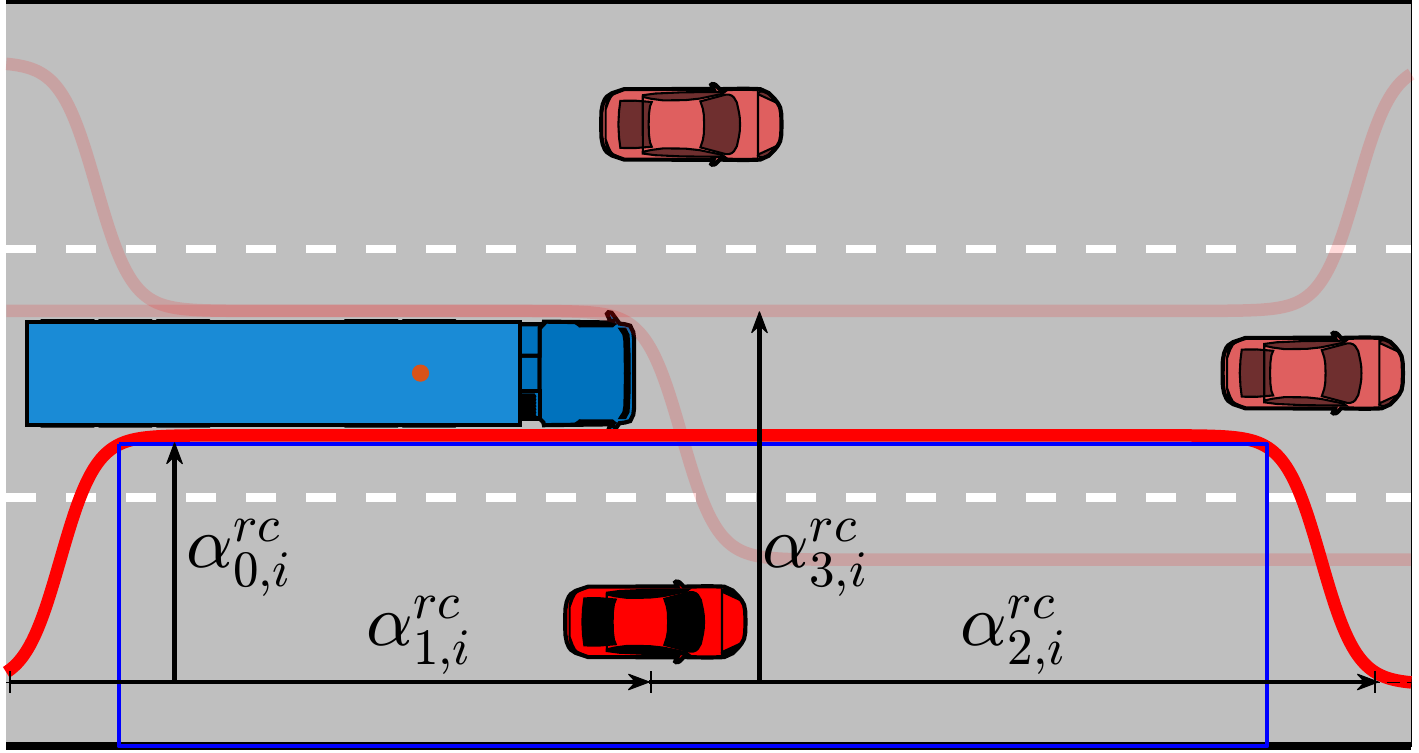}
    \caption{Collision avoidance constraints for the right lane change controller.}
    \label{fig:constraint_changeright}
    \end{subfigure}
    \caption{Collision avoidance constraints for each controller. Red vehicles are explicitly included in constraints, indicated metrics refer to the opaque vehicles and constraints. The constrained point on the ego-vehicle is indicated with a red dot, (b) and (c) additionally display the approximated discontinuous constraint.}
    \label{fig:constraints}
\end{figure*}

For the lane keeping controller $j = \tx{nc}$, avoiding collisions amounts to maintaining a longitudinal distance to a leading vehicle $i$, while remaining in the current lane. The minimum allowed longitudinal distance has to satisfy
\begin{equation}\small \label{eq:nc_constraint}
\begin{split} 
    c_\tx{e}^{\tx{nc}}(\xx_\tx{e}^{\tx{nc}},\hat{\xx}_i^\tx{nc}) & = p_{x,\tx{e}}^{\tx{nc}} - \hat{p}^\tx{nc}_{x,i} + \ell_1 + d_s + T_s v_{x,\tx{e}}^{\tx{nc}} + \lambda^\tx{nc}_{\tx{e},i} < 0
\end{split}
\end{equation}%
where $d_s$ represents a distance margin, $\ell_1$ is the tractor length, $T_s$ represents the time-headway and $\lambda^\tx{nc}_{\tx{e},i}$ is the slack variable. To avoid collisions with vehicles in adjacent lanes, the lateral position is constrained to the current lane margins, accounting for the maximum lateral span of the ego-vehicle $d_\tx{w,e}$.
A visualization of the constraints with the referenced parameters is displayed in Fig. \ref{fig:constraint_keeplane}.

In the lane changing case, $j \in \{\tx{lc},\tx{rc} \}$, we aim to constrain both the longitudinal and lateral distance to some vehicle $i$. This yields a possibly non-differentiable constraint which can be formulated as a mixed integer problem (MIP). However, as MIPs are significantly more computationally intensive to solve, we employ a smooth approximation $\Tilde{y}_i^j$ of the constraint boundaries using $\tanh$-functions, as
\begin{equation} \small
    \begin{split}
    \Tilde{y}_i^j(\xx_\tx{e}^j,\hat{\xx}^j_i) &= \frac{\alpha_{0,i}^j (\hat{p}^j_{y,i})}{2} \Big (  \tanh{\big ( p_{x,\tx{e}}^j - \hat{p}^j_{x,i} + \alpha_{1,i}^j \big )}\\
    &+\tanh{\big (\hat{p}^j_{x,i} - p_{x,\tx{e}}^j + \alpha_{2,i}^j \big )} \Big ) + \alpha_{3,i}^j(p_{y,\tx{e}}^j,\hat{p}^j_{y,i}).
    \end{split}
\end{equation}
Here, $\alpha_{0,i}^j$ and $\alpha_{1,i}^j$ scale the constraint in lateral and longitudinal distance, respectively, and incorporate the surrounding vehicle position, width and length, together with those of the ego-vehicle, plus a 
safety margin. 
Lastly, $\alpha_{3,i}^j$ shifts the constraint in lateral distance based on the current lateral position of the surrounding and ego-vehicle. A visualization of the constraints for both left and right lane changes are displayed in Fig. \ref{fig:constraint_changeleft} and Fig. \ref{fig:constraint_changeright}. 
Hence, the collision avoidance constraints can be formulated as,
\begin{equation} \label{eq:lc_rc_constraint} \small
    c_\tx{e}^{j}(\xx_\tx{e}^j,\hat{\xx}^j_i) = \beta_i^j \big (\Tilde{y}_i^j(\xx_\tx{e}^j,\hat{\xx}^j_i)+d_\tx{w,e} - p_{y,\tx{e}}^j + \lambda^j_{\tx{e,i}} \big ) < 0
\end{equation}
where $\beta_i^j \in \{-1,1\}$ converts between an upper and lower constraint on lateral position, depending on the vehicle $i$ and controller $j$. This constraint is repeated for all surrounding vehicles $i \in \mathbb{M}^j_\tx{v}$.

Finally, constraints \eqref{eq:nc_constraint} and \eqref{eq:lc_rc_constraint} together with the physical limitations on the ego-vehicle states and control input form the general constraints for each controller $j$,
\begin{equation} \label{eq:ineq_eqo}
    h_\tx{e}^j(\mathbf{x}_\tx{e}^j(k),\mathbf{u}_\tx{e}^j(k),\lam_{\tx{e}}^j(k),\hat{\xx}^j(k),\hat{\uu}^j(k)) \leq 0,
\end{equation}
where $\hat{\xx}^j(k)$ and $\hat{\uu}^j(k)$ gather the estimated/predicted states and control inputs of all surrounding vehicles. 

\subsection{Optimal Control Problem Formulation}
Each MPC $j$ of the ego-vehicle aims to track a reference in longitudinal velocity and lateral position while minimizing the control input. The objective also addresses driver comfort by minimizing the change in acceleration and steering angle. Expressing the objective over the horizon for each controller $j$, including slack variables, yields, 
{\allowdisplaybreaks
\begin{subequations}\label{eq:ego_objective}\small
    \begin{align} \label{eq:ego_obj_x_f}
    f_\tx{e}^j(\XX_\tx{e}^j, \UU_\tx{e}^j,&\LL_\tx{e}^j)= \; ||\xx_\tx{e}^j(N)-\xx_{\tx{e},\tx{r}}^j(N)||^2_P +||\lam_\tx{e}^j(N)||^2_{Q_s}\\ \label{eq:ego_obj_x}
    &+\sum_{k=0}^{N-1} ||\xx_\tx{e}^j(k)-\xx_{\tx{e},\tx{r}}^j(k)||^2_Q+ ||\lam_\tx{e}^j(k)||^2_{Q_s}\\
    &+\sum_{k=0}^{N-1} ||\uu_\tx{e}^j(k) - \uu_{\tx{e}, \tx{r}}^j(k)||^2_R\\ \label{eq:ego_obj_comfort}
    &+\sum_{k=0}^{N-2} ||\uu_\tx{e}^j(k+1) - \uu_\tx{e}^j(k)||^2_{R_d}
    \end{align}
\end{subequations}}%
where ${Q_s = q_s I_{{M}_c}}$, ${R = \text{diag}(q_a,q_\delta)}$, 
${R_d = \text{diag}(q_{da},q_{d\delta})}$ and ${Q = \text{diag}(0,q_y,q_v,0,0)}$ are positive semi-definite matrices and ${\xx_{\tx{e},\tx{r}}^j = [0,p_{y,\tx{r}}^j,v_{x,\tx{r}},0,0]^T}$, ${\uu_{\tx{e},\tx{r}}^j = [0,0]^T}$ represent the state and input reference for each controller. The ego vehicle states over the horizon are gathered in 
\begin{align} \label{eq:ego-state-traj}
    \XX_\tx{e}^j &= \{ \xx_\tx{e}^j(0), \xx_\tx{e}^j(1), \dots, \xx_\tx{e}^j(N) \}, \quad j \in \{\tx{nc}, \tx{lc}, \tx{rc}\}
\end{align}
and similarly are the control inputs and slack variables in $\UU_\tx{e}^j$ and $\LL_\tx{e}^j$.
The positive definite terminal cost matrix $P$ is determined using an infinite horizon linear quadratic regulator (LQR) by solving the discrete time Riccati differential equation for linearized dynamics \eqref{eq:ego_dynamics} around $\xx_{\tx{e},\tx{r}}$ and $\uu_{\tx{e},\tx{r}}$, see \cite{rawlings} for more details.

We can now express the optimal control problem for the ego-vehicle and each respective controller $j \in \{ \tx{nc},\tx{lc},\tx{rc}\}$,
{\allowdisplaybreaks
\begin{subequations} \label{eq:P_ego}
\begin{align}
         \min_{\XX_\tx{e}^j,\UU_\tx{e}^j,\LL_\tx{e}^j} & f_\tx{e}^j(\XX_\tx{e}^j,\UU_\tx{e}^j,\LL_\tx{e}^j)\\ \label{eq:P_ego_eq}
        \text{s.t. } &\mathbf{x}_\tx{e}^j(k+1) = g_\tx{e}(\mathbf{x}_\tx{e}^j(k),\mathbf{u}^j_\tx{e}(k))\\\label{eq:P_ego_ineq_ego}
        & h_\tx{e}^j(\mathbf{x}^j_\tx{e}(k),\mathbf{u}^j_\tx{e}(k),\lam_{\tx{e}}^j(k),\hat{\xx}^j(k),\hat{\uu}^j(k)) \leq 0 \\ \label{eq:P_ego_ineq_slack}
        &\lam_{\tx{e}}^j(k) \leq 0\\
        &\mathbf{x}_\tx{e}^j(0) = \mathbf{x}_{0,e} \label{eq:P_ego_init_state}\\
        & (\hat{\XX}^j,\hat{\UU}^j) = \pppi(\cdot, \XX_\tx{e}^j,\UU_\tx{e}^j) \label{eq:P_ego_predictor}
\end{align}
\end{subequations}}%
where \eqref{eq:P_ego_eq}--\eqref{eq:P_ego_ineq_slack} are imposed $\forall k = 0, \dots ,N-1$ and $(\hat{\XX}^j,\hat{\UU}^j)$ are estimated/predicted trajectories of surrounding vehicles over the horizon, gathering $\hat\xx^j(k)$ and $\hat\uu^j(k)$, $\forall k$ similarly as in \eqref{eq:ego-state-traj}.
The predictor $\pppi$ will be described in more details in Section~\ref{sec:PP-DMPC}. It is important to note here that sensitivities (derivatives) of the predictor $\pppi$ may not be available and problem \eqref{eq:P_ego} may not be possible to directly cast to a standard nonlinear program (NLP). Indeed, the goal of this paper is to decouple \eqref{eq:P_ego_predictor} such that the subproblem in \eqref{eq:P_ego} can be cast as a standard NLP. 

After each of the three MPCs are finished computing, the decision manager selects the control signals from the MPC that optimizes,
\begin{equation} \label{eq:decision-manager}
    \argmin_{\XX_\tx{e}^j,\UU_\tx{e}^j} \; q_\tx{e} f_\tx{e}^j(\cdot) + q_\tx{c} f_\tx{c}^j + q_\tx{s} f_\tx{s}^j(\xx_{0,e}), \;j \in \{\tx{nc},\tx{lc},\tx{rc}\}
\end{equation}
where $q_\tx{e},q_\tx{c},q_\tx{s}$ are positive scaling parameters and $f_\tx{e}^j$ is current cost of the $j^\text{th}$ MPC. The function $f_\tx{c}^j$ introduces a cost on switching $j$ based on $m$ previous decisions as,
\begin{equation}
    f_\tx{c}^j = \sum_{i=1}^m \indicator\left(H(i)-j \right)
\end{equation}
 where $H$ is the history of previous choices of $j$ and $\indicator$ is an indicator function. The function $f_\tx{s}^j$ introduces a cost on the distance to a highway exit as,
\begin{equation}
    f_\tx{s}^j(\xx_{0,e}) = \left ( 1 - \left( \frac{d_\tx{exit} - p_{0,x}}{d_\tx{max}}\right)^\gamma \right) \indicator( j^\star - j)
\end{equation}
where $d_\tx{exit}$ is the distance to the exit, $d_\tx{max}>d_\tx{exit}$, $\gamma\in[0,1]$ are considered tuning parameters and $j^\star$ is the decision that guides the ego-vehicle towards the highway exit\cite{karlssonExitMPC}.




\section{Predicting and Planning using DMPC}\label{sec:PP-DMPC}
As outlined in Section \ref{sec:ego_problem}, a trajectory planner in highway traffic can be designed using MPC with local objectives and constraints, combined with coupled inequality constraints. Without vehicle-to-vehicle communication, the coupled constraints require a prediction of the surrounding vehicles trajectories, which in turn depend on the planned ego-vehicle trajectories.
Consider a general predictor $\pppi$ as in \cite{trajectron++}, that returns trajectories $(\hat{\XX},\hat{\UU})$ over a future horizon. The predictor utilizes past observations of all vehicles $\XX_{\tx{obs}},\UU_{\tx{obs}}$ together with the planned trajectory of the ego-vehicle $\XX_{\tx{e}},\UU_{\tx{e}}$ over the future horizon as,
\begin{equation}
    {(\hat{\XX},\hat{\UU}) = \pppi(\XX_{\tx{obs}},\UU_{\tx{obs}},\XX_\tx{e},\UU_\tx{e}).}
\end{equation}
In the DMPC formalism, the predicted variables can be treated as an approximation of the local solution of each agent's optimal control problem \eqref{eq:P_dist}. Hence, if the prediction $(\hat{\XX}^p,\hat{\UU}^p)$ at some iterate $p$ can be used to express the coupled constraints, it is possible to apply the method presented in Section \ref{sec:DMPC_theory} to formulate an iterative algorithm that attempts to solve the centralized problem by distributing the ego-vehicle planner from its predictor. Given a fixed prediction $(\hat\XX^p,\hat\UU^p)$, problem \eqref{eq:P_ego} becomes a smooth NLP as,
\begin{subequations}\label{eq:P_ego_subproblem}
\begin{align}
         P_\tx{MPC}(\hat\XX^p,\hat\UU^p)=&\argmin_{\XX_\tx{e},\UU_\tx{e},\LL_\tx{e}}f_\tx{e}(\XX_\tx{e},\UU_\tx{e},\LL_\tx{e}) \\
         &\text{ s.t. } \eqref{eq:P_ego_eq} - \eqref{eq:P_ego_init_state}%
\end{align}
\end{subequations}
here, and onward omitting the index $j$ for convenience.
Predicted and planned trajectories that remain constant over iterates $p$ indicate that both concur. 
Note that the convergence that has been guarantied for linear systems in \cite{rawlings} cannot be generalized to this problem as we consider non-linear MPC with prediction errors and a general predictor. To prevent divergence, we propose a loss metric $L$, which can treat prediction and planning jointly,
\begin{equation} \label{eq:loss-function}
\begin{split}
       {L^{p+1}} &= ||\hat{\XX}^{p+1}-\hat{\XX}^{p}||_2 + ||\hat{\UU}^{p+1}-\hat{\UU}^{p}||_2\\
       &+ ||\XX_\tx{e}^{p+1}-\XX_\tx{e}^{p}||_2 + ||\UU_\tx{e}^{p+1}-\UU_\tx{e}^{p}||_2.
\end{split}
\end{equation}
Our approach enforces the loss metric to decrease at each iteration, i.e $L^{p+1} < L^{p}$. If the loss metric stops decreasing, or if it drops below a tolerance $\epsilon$, the iteration is terminated. This trivially means that the loss metric cannot diverge from the closest local solution of the trajectory initialization. 

\begin{algorithm}[ht] 
\caption{Predicting and Planning using DMPC}
\begin{algorithmic}[1]
 \FOR{ $\kappa = 1, \dots, \kappa_\tx{end}$}
 \STATE Initialize: $(\XX^0_{\tx{e},\kappa},\UU_{\tx{e},\kappa}^0)\! = \!\shift (\XX_{\tx{e},\kappa-1}^\star,\UU_{\tx{e},\kappa-1}^\star) $\\
 \hspace{1.5cm}$(\hat\XX^0_{\kappa},\hat\UU_{\kappa}^0)\! =\! \pppi(\XX_{\tx{obs},\kappa},\UU_{\tx{obs},\kappa}, {\XX}_{\tx{e},\kappa}^0{\UU}_{\tx{e},\kappa}^0)$\\
    \hspace{1.5cm}$L^0 = \infty$
 \FOR{$p = 0,\dots, p_\tx{max}$} 
     \STATE Plan: $(\XX^{p+1}_{\tx{e},\kappa}, \UU^{p+1}_{\tx{e},\kappa}) ^\star = P_\tx{MPC}(\hat{\XX}_\kappa^{p},\hat{\UU}_\kappa^{p})$ \label{alg:plan}\\
    \STATE Update: ${\XX}_{\tx{e},\kappa}^{p+1} \leftarrow w_\tx{e} ({\XX}_{\tx{e},\kappa}^{p+1})^\star + (1-w_\tx{e}) {\XX}_{\tx{e},\kappa}^p$ \label{alg:update-ego}
    
    \hspace{1.1cm} ${\UU}_{\tx{e},\kappa}^{p+1} \leftarrow w_\tx{e} ({\UU}_{\tx{e},\kappa}^{p+1})^\star + (1-w_\tx{e}) {\UU}_{\tx{e},\kappa}^p$
    \STATE Predict:     ${(\hat{\XX}_\kappa,\hat{\UU}_\kappa) \!= \! \pppi(\XX_{\tx{obs},\kappa},\UU_{\tx{obs},\kappa}, {\XX}_{\tx{e},\kappa}^{p+1},{\UU}_{\tx{e},\kappa}^{p+1})}$
    \STATE Update: $\hat{\XX}_\kappa^{p+1} \leftarrow w \hat{\XX}_\kappa + (1-w) \hat{\XX}_\kappa^p$
    
    \hspace{1.1cm} $\hat{\UU}_\kappa^{p+1} \leftarrow w \hat{\UU}_\kappa + (1-w) \hat{\UU}_\kappa^p$
    \STATE Compute: $L^{p+1}$ using \eqref{eq:loss-function}
        \IF{ $L^{p} < L^{p+1}$}
    \STATE 
    $\big(\XX_{\tx{e},\kappa}^\star, \UU_{\tx{e},\kappa}^\star \big) = (\XX^{p}_{\tx{e},\kappa}, \UU^{p}_{\tx{e},\kappa}) ^\star$, exit the loop
    \ELSIF{ $L^{p+1} < \epsilon$}
    \STATE 
    $\big(\XX_{\tx{e},\kappa}^\star, \UU_{\tx{e},\kappa}^\star \big) \! = (\XX^{p+1}_{\tx{e},\kappa}, \UU^{p+1}_{\tx{e},\kappa}) ^\star$, exit the loop
    \ENDIF
 \ENDFOR
 \STATE Apply the first control action $u_\tx{e}(0|\kappa)$ from $\UU_{\tx{e},\kappa}^\star$
 \ENDFOR
\end{algorithmic}
\label{alg:DMPC}
\end{algorithm}
 Algorithm \ref{alg:DMPC} displays the formulation of the coupled predicting and planning method, employing the DMPC scheme for \eqref{eq:P_ego_subproblem}. In a proper MPC setting, we introduce an index $\kappa$ when MPC is updated, such that each variable, e.g., the ego vehicle state, can be written as $\xx_\tx{e}(\kappa+k|\kappa)$, or even simpler as $\xx_\tx{e}(k|\kappa)$. The predicted trajectories over the horizon, e.g., the states as in \eqref{eq:ego-state-traj}, can be written as $\XX_{\tx{e},\kappa}^p$. In line~\ref{alg:plan}, the algorithm computes an optimal MPC solution, denoted as $(\XX^{p+1}_{\tx{e},\kappa}, \UU^{p+1}_{\tx{e},\kappa})^\star$, by iterating the DMPC method over $p$ towards $p_\tx{max}$ at each time step $\kappa$ until $\kappa_\tx{end}$. 
 Notice that the ego trajectories updated in line~\ref{alg:update-ego} are used only for computing the loss metric and by the predictor, while the returned DMPC solution $(\XX_\tx{e,\kappa}^*, \UU_\tx{e,\kappa}^*)$ is from one of the iterates in line~\ref{alg:plan}, i.e., before performing the update.

 To derive a good initial guess for variables at time step $\kappa$, we define a function $\shift(\cdot)$ that shifts the optimal variables from time step $\kappa-1$ and pads the last value  using zero control\cite{rawlings}.


\section{Simulation Study} \label{sec:results}
This section provides a case study with the proposed methods, investigating performance and convergence with respect to prediction errors.

\subsection{Simulation Setup}
The simulation environment\footnote{Code: https://github.com/BorveErik/Autonomous-Truck-Sim} controls the surrounding vehicles using the models described in Section \ref{sec:traffic_model} and the ego-vehicle uses our MPC. The traffic model parameters are randomized using uniform distributions as ${\boldsymbol{\phi} = \boldsymbol{\phi}_\tx{mean} + \mathcal{U}(\boldsymbol{\phi}_\tx{min},\boldsymbol{\phi}_\tx{max})}$. 
The MPC problem is formulated using \textit{Casadi}, running the \textit{IpOpt} solver \cite{casadi}. We demonstrate the approach presented in Section \ref{sec:PP-DMPC} using a model-based predictor $\pppi$, utilizing the model in Section \ref{sec:traffic_model} with added noise $\mathcal{N}(0,\sigma_\tx{a})$ to the control inputs of the surrounding vehicles. 

The predictor hence utilize a measurement of surrounding vehicles, repeatedly applies the ground-truth model using the planned ego-vehicle trajectory and adds normal distributed noise to introduce a prediction error that can be scaled with $\sigma_\tx{a}$. 
Note that this predictor will be less ideal in estimating and predicting the motion of real vehicles, which may require learning-based methods. Yet, we deem this approach suitable to clearly demonstrate the proposed DMPC algorithm and investigate the impact of prediction errors. 

\begin{table*}[ht]
    \vspace{4mm}
    \centering
    \caption{Statistic results for all 100 sampled scenarios and each predictor as ${\tiny(\sigma_{\tx{a},1}/\sigma_{\tx{a},2}/\sigma_{\tx{a},3})}$. Relative controller cost are presented as a percentage.}
    \begin{tabular}{|c||c|c|c||c||c|c|}\hline
         Controller & Success (\%) & Collision (\%) & Time (s) & Total Cost (\%)& Avg. Nr. of Iterations ($p$) & Convergence Rate (\%)\\ \hline \hline
         DC-MPC & 100/99/98 & 0/1/2 & 22.1/22.1/22.1 & 97.3/98.1/100.0  & --- & ---\\\hline
         PP-DMPC  & 100/100/99 & 0/0/1 & 21.1/21.2/21.5 & 66.0/71.8/74.1 & 2.41/2.68/2.38 & 87.0/64.1/17.4\\\hline
    \end{tabular}
    \label{tab:results}
\end{table*}

\subsection{Test Scenarios}
The dense traffic environment is simulated by randomly sampling $M$ vehicles in close proximity of the ego-vehicle located in the middle lane. Longitudinal inter-vehicle distances in adjacent lanes are kept smaller than the length of the ego-vehicle. The simulations are initialized such that a lane change is feasible, if properly handled by the controller. Fig. \ref{fig:traj_plan} illustrates a sampled scenario. In this setting, we consider a \textit{``Forced lane change''} maneuver (FLC) where the ego-vehicle is required to reach a highway exit at ${p_{x,\tx{exit}} = \SI{250}{\meter}}$, ahead in the rightmost lane. The desired speed $v_{x,\tx{r}}$ is \SI{30}{\kilo\meter\per\hour} and a lane change is defined as successful if the right lane's center $p_{y,\tx{r}}^{rc}$ is reached before the exit. The maximum considered time frame is ${t_\tx{max} = \SI{30}{\second}}$. We study two separate controller architectures for the ego-vehicle: 1.) \textit{``DC-MPC''}: The ego-vehicle considers decoupled prediction and planning; 2.)\textit{``PP-DMPC''} The ego-vehicle couples prediction and planning using Algorithm~\ref{alg:DMPC}. The PP-DMPC hyperparameters are chosen as: $p_\tx{max} = 15$, $w=w_\tx{e} = 1/(M+1)$ and $\epsilon = 5$. Hence, both controllers utilize the same predictive capabilities of $\pppi$ and gauge interactions with other vehicles by first approaching the lane markers of the desired lane. The PP-DMPC controller further considers how the ego-vehicle trajectory interacts with the predicted trajectory of the surrounding vehicles through Algorithm \ref{alg:DMPC}.
Simulations are evaluated with three different noise settings for $\pppi$: ${\sigma_{\tx{a},1}= 0.1}$, ${\sigma_{\tx{a},2}= 0.5}$, and ${\sigma_{\tx{a},3}= 1.0}$. Note that the acceleration of each vehicle is confined within a defined physical limitation of $\pm$\SI{4}{\meter\per\second\squared}, both in simulation and in the predictor.

\subsection{MPC Performance Results}
\subsubsection{Qualitative Analysis}
\begin{figure}[t!]
    \vspace{1mm}
    \centering
    \begin{subfigure}{1 \textwidth}
    \includegraphics[width = 0.48 \textwidth]{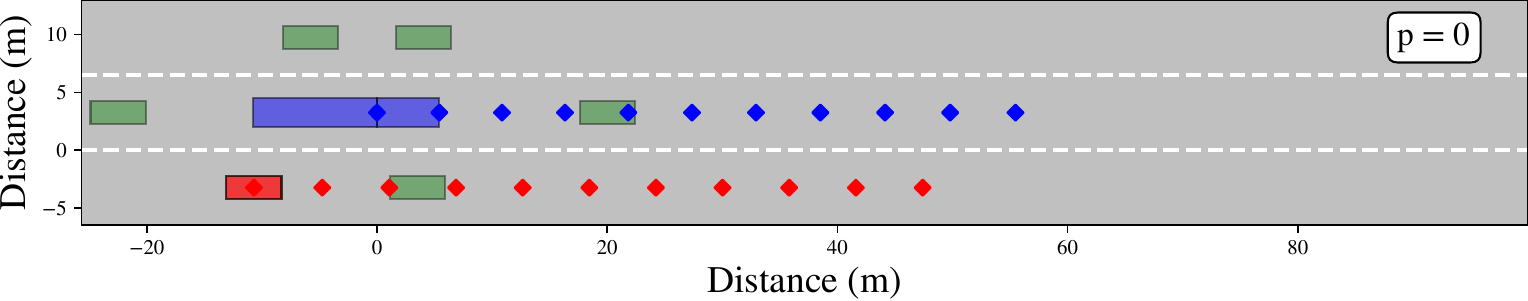}
    \end{subfigure}
    \begin{subfigure}{1 \textwidth}
    \includegraphics[width = 0.48 \textwidth]{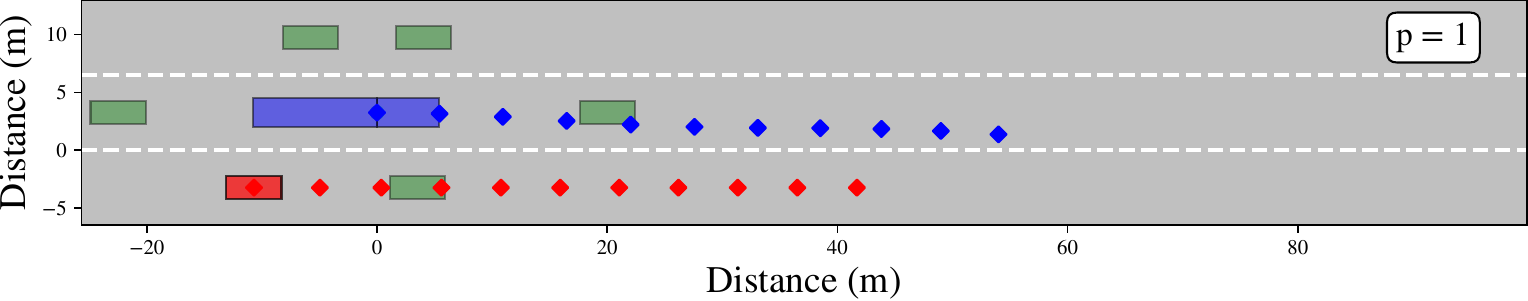}
    \end{subfigure}
    \begin{subfigure}{1 \textwidth}
    \includegraphics[width = 0.48 \textwidth]{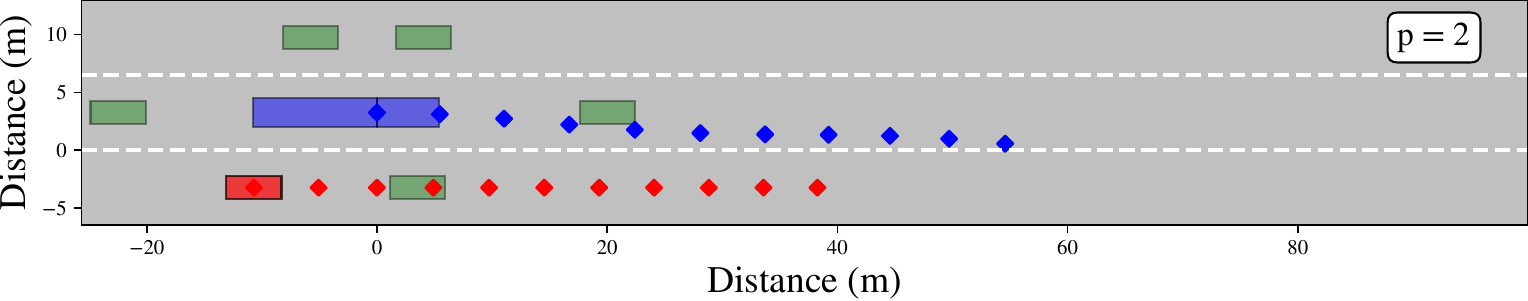}
    \end{subfigure}
    \caption{Example of a generated FLC scenario. Each plot displays the state trajectory of two vehicles in corresponding color at, iteration $p$.}
    \label{fig:traj_plan}
\end{figure}

Fig. \ref{fig:traj_plan} highlights the key characteristics of the coupled trajectory prediction and planning proposed in our method. The top plot displays a naive, but feasible initialization of 
the state trajectories for the ego-vehicle and an adjacent vehicle. 
As iterations in the distributed scheme continue, the predictor and planner reason via coupled variables. The middle plot displays emerging interactions between the planned ego-vehicle trajectory (blue) and an adjacent vehicle (purple) at $p=5$. Finally, the bottom plot displays a converged PP-DMPC solution at $p=15$, where the adjacent vehicle in the right lane have slowed down to let the ego vehicle make a lane change. 

\subsubsection{Quantitative Analysis}
To evaluate the performance of both controllers, 100 scenarios were sampled. The overall performance is evaluated using the following: 1.) \textit{Success Rate}'': The percentage of scenarios that achieve the goal of the respective case; 2.) ``Collision rate'': The percentage of scenarios resulting in a collision; 3.) ``Time'': The average completion time for the successful scenarios.
We further treat ``optimality'' by considering: ``Total Cost'' as an evaluation of the objective \eqref{eq:ego_objective} for all simulated trajectories. These metrics are presented in Table \ref{tab:results} where results for different predictors are indicated as $\sigma_{\tx{a},1}/\sigma_{\tx{a},2}/\sigma_{\tx{a},3}$. The PP-DMPC out-performs the DC-MPC in the investigated metrics, for all different noise levels. The performance of both the DC-MPC and the PP-DMPC decreases with an increasing $\sigma_\tx{a}$. The relative improvements of the PP-DMPC over the DC-MPC also decreases with an increasing $\sigma_\tx{a}$. This indicates that Algorithm \ref{alg:DMPC}, applied for our non-linear MPC can have benefits, however also that the extent of which is reliant on the predictors accuracy.

\subsection{Convergence and Prediction Errors}
\begin{figure}
    \centering
    \includegraphics[width = 0.48 \textwidth,trim={0 0 0 0},clip]{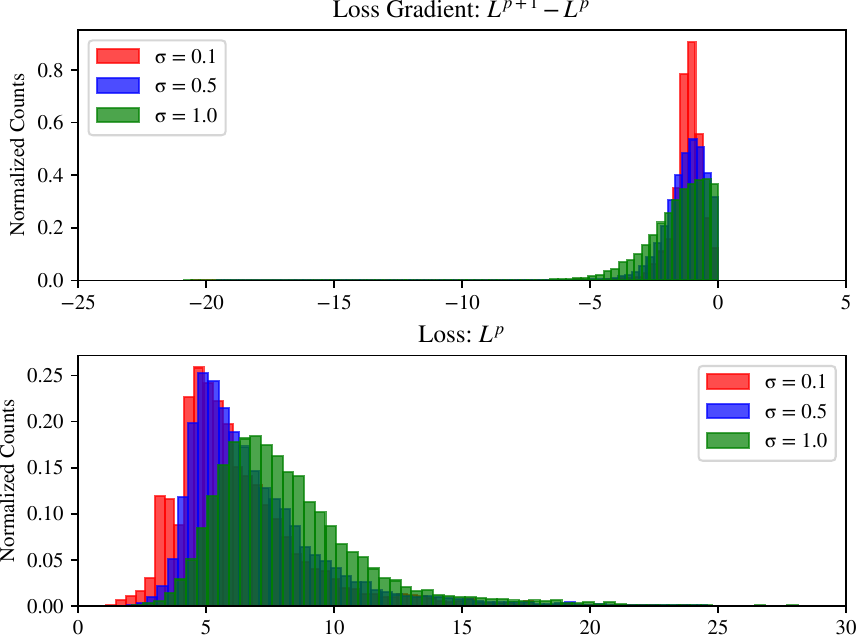}
    \caption{Distribution of the loss (bottom) and its gradient (top).}
    \label{fig:loss}
\end{figure}

The average number of iterations for the PP-DMPC and the convergence rate for each predictor is displayed in Table \ref{tab:results}. This indicates that the convergence of Algorithm \ref{alg:DMPC} is dependent on the accuracy of the predictor. The distribution of the loss $L$ and its gradient $L^{p+1}-L^p$, is displayed in Fig. \ref{fig:loss}.  On average, larger prediction errors result in a larger loss and larger corresponding gradients. The loss gradients are indeed never positive which empirically shows that the algorithm does not diverge from the closest local minima.

\section{Conclusions and Future Work}
The largest challenge for our proposed method is related to real-world applicability. Based on our simulation study, we hope to investigate if learning-based predictors, trained on real-world traffic data, can obtain a sufficient accuracy for our proposed method. Another challenge relates to the computational complexity. We hope to investigate this in future work, for example by considering a sub-optimal MPC formulation. Our constraint on a continual decreasing loss is also conservative and might limit exploration of the non-linear state space. A more rigorous study of convergence might reveal more effective criteria and ideal hyper-parameter choices, which could improve MPC performance and the convergence rate. Lastly, as the collision avoidance constraints need to be relaxed to aid feasibility, the planned trajectory can result in collisions in edge cases. This could be further mitigated by several approaches e.g., considering robust MPC \cite{trajectoryReview} with stochastic predictors \cite{trajectron++}, and identifying and aborting dangerous maneuvers \cite{nilssonSimulator}.

\section*{Acknowledgment}
The authors would like to thank Morteza Haghir Chehreghani, Deepthi Pathare and Stefan Börjesson for our valuable discussions.

\addtolength{\textheight}{-12cm}   





\bibliographystyle{IEEEtran}
\bibliography{IEEEabrv,main.bib}

\end{document}